\documentclass{llncs}
\usepackage{latexsym,amssymb,upgreek,graphicx}
\usepackage{enumerate}
\usepackage{amsmath}
\usepackage{amsfonts}
\usepackage{xspace}
\usepackage{color}
\usepackage{url}
\usepackage{mathrsfs}
\usepackage{listings}
\usepackage{enumitem}
\usepackage{hyperref}

\newcommand{\bpdww}{Bit-Parallel$^2$ Wide-Window\xspace}
\newcommand{\bpwwd}{Bit-Parallel Wide-Window$^2$\xspace}

\newcommand{\p}{p\xspace}

\newcommand{\bigO}{\mathcal{O}}

\newcommand{\myitem}[4]{%
  \item{	 
  		\label{#1}
 		{\textbf{#2}} (\small\textsc{#1}) - #4\\
  		Appeared in #3\\
	}
} 

\newcommand{\mw}{m/w}

\begin{document}

\title{Exact Online String Matching Bibliography}

\titlerunning{Exact Online String Matching Bibliography}

\author{Simone Faro}

\institute{Universit\`{a} degli Studi di Catania, Dipartimento di Matematica e Informatica\\
Viale Andrea Doria 6, I-95125, Catania, Italy\\
\email{faro@dmi.unict.it}
}

\maketitle

\begin{abstract}
In this short note we present a comprehensive bibliography for the \emph{online exact string matching problem}. 
The problem consists in finding \emph{all} occurrences of a given pattern in a text.
It is an extensively studied problem in computer science, mainly due to its direct applications to such diverse areas as text, 
image and signal processing, speech analysis and recognition, data compression, information retrieval, computational biology and chemistry.
Since 1970 more than 120 string matching algorithms have been proposed.
In this note we present a comprehensive list of (almost) all string matching algorithms. The list is updated to May 2016.
\end{abstract}

\markboth{Simone Faro}{Exact Online String Matching Bibliography}

\section{Introduction}\label{intro}
Given a text $t$ of length $n$ and a pattern $\p$ of length $m$ over some alphabet $\Sigma$ of size $\sigma$, 
the \emph{string matching problem} consists in finding \emph{all} occurrences of the pattern $\p$ in the text $t$.  
String matching algorithms are also basic components used in implementations
of practical softwares existing under most operating systems. Moreover, they emphasize programming methods that serve as paradigms in other fields
of computer science. Finally they also play an important role in theoretical computer science by providing challenging problems.
Applications require two kinds of solutions depending on which string, the pattern or the text, is given first.
Algorithms based on the use of automata or combinatorial properties of strings are commonly implemented to preprocess the
pattern and solve the first kind of problem. This kind of problem is generally referred as \emph{online} string matching. 
The notion of indexes realized by trees or automata is used instead in the second kind of problem, generally referred as \emph{offline} string matching. In this note we are only interested in algorithms of the first kind.

Since 1970 more than 120 online string matching algorithms have been proposed.
All solutions can be divided into several different classes. Here we divide all algorithms into four classes, although different classifications are possible.
Specifically we distinguish algorithms which solve the problem by making use of comparisons between characters, algorithms based on packed string matching, and algorithms which make use of automata. 
The latter class can be further divided into two classes:
algorithms which make use of deterministic automata and algorithms based on bit-parallelism which simulate the behavior of non-deterministic automata.

\begin{figure}
\begin{center}
\includegraphics[width=0.99\textwidth]{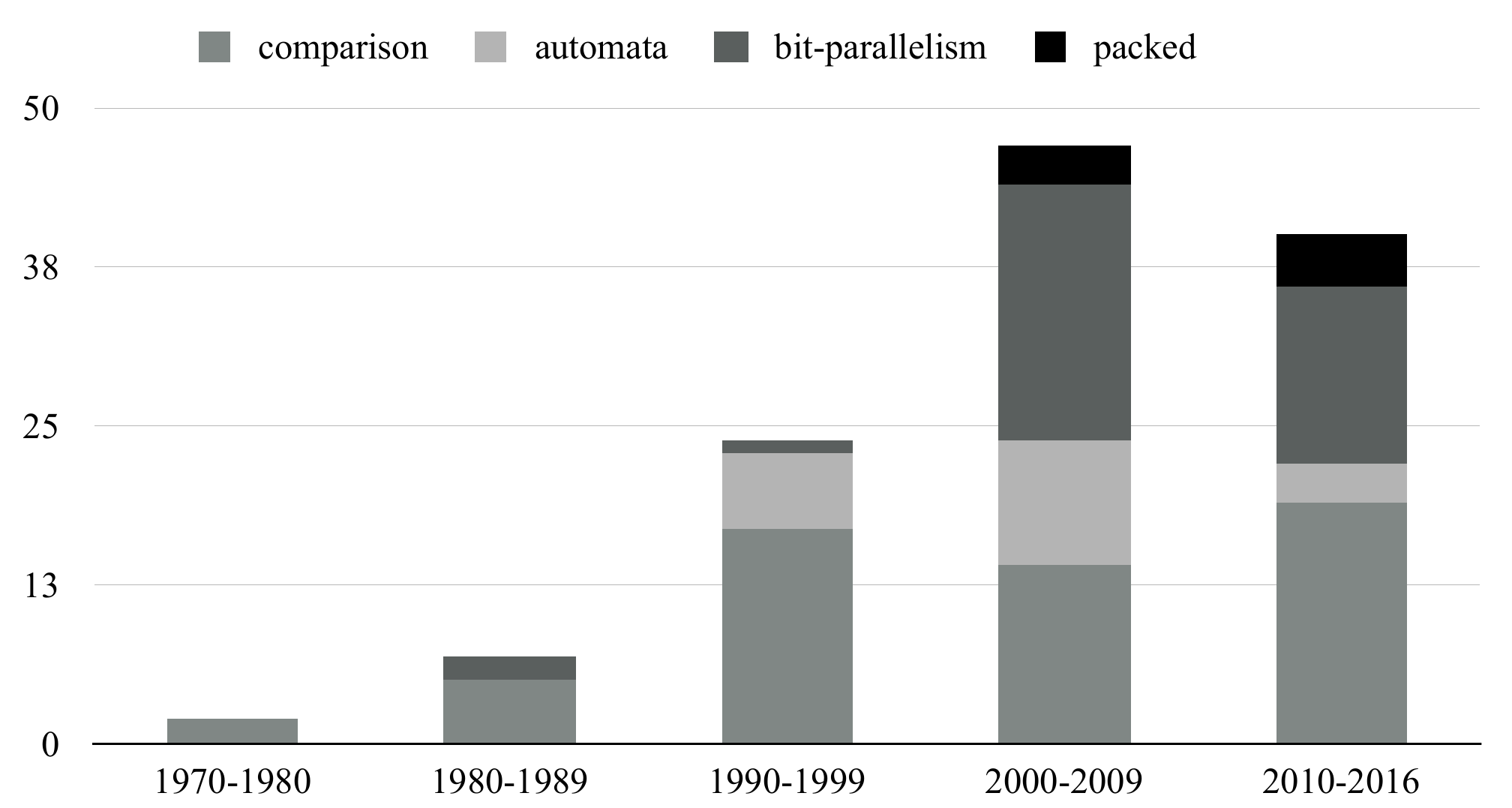}
\end{center}
\caption{Number of appeared string matching algorithms since 1970.}
\end{figure}

The lists of string matching algorithms proposed in the following sections are in chronological ordered. Each entry of the lists includes the name of the algorithm, a small acronym used in literature to refer to the algorithm, the list of references to the paper where the algorithm appeared and a very short description of the main features of the algorithm.
The Smart tool\footnote{Smart (String Matching Algorithms Research Tool), by Simone Faro and Thierry Lecroq, is available at \url{http://www.dmi.unict.it/~faro/smart/}} includes the implementation of (almost) all the algorithms presented in this note. The small acronym associated to each algorithm is used in Smart to refer to the algorithm.

\def\labelitemi{\tiny$\blacksquare$}

\section*{List of Algorithms based on Comparison of Characters}
In a computational model where the matching algorithm is restricted to read all the characters of the text one by one the
optimal complexity is $\bigO(n)$, and was achieved the first time by the well known Morris-Pratt algorithm
\cite{MP70}.
However, in many practical cases it is possible to avoid reading all the characters of the text achieving sub-linear
performances on the average. 

The optimal average $\bigO(\frac{n\log_{\sigma}m}{m})$ time complexity \cite{Yao79} was reached by many comparison based algorithms.  
However, all algorithms with a sub-linear average behavior may have to read all the text characters in the worst
case. 
It is interesting to note that many of those algorithms have an even worse $\bigO(nm)$-time complexity in the
worst-case \cite{FL10,FL13}.

Int what follows we present the list of all string matching algorithms based on comparison of characters.\\[0.2cm]

\begin{enumerate}
\itemsep0.2cm
\myitem{{BF}} {Brute-Force} {~\cite{CLRS01}} {no date} 
	{Naive algorithm for the string matching problem. It works in $\bigO(nm)$-time}
\myitem{{MP}} {Morris-Pratt} {~\cite{MP70}} {1970} 
	{First linear algorithm scanning the text and the pattern from left to right. It derives from (\ref{{BF}}). Preprocessing of the pattern in $\bigO(m)$-time and -space.}
\myitem{{KMP}} {Knuth-Morris-Pratt} {~\cite{KMP77}} {1977} 
	{Linear algorithm. Improvement of (\ref{{MP}}).}
\myitem{{BM}} {Boyer-Moore} {~\cite{BM77}} {1977} 
	{Introduces the sliding window approach and the scanning of the window from right to left. It works in $\bigO(nm)$-time. 
	It uses the occurrence and the good-suffix heuristics.
	It derives from (\ref{{BF}})}
\myitem{{HOR}} {Horspool} {~\cite{Hor80}} {1980} 
	{Improvement of (\ref{{BM}}). It uses only a modification if the occurrence heuristics.}
\myitem{{GS}} {Galil-Seiferas} {~\cite{GS81,GS83}} {1981} 
	{Linear algorithm using constant extra space complexity. Preprocessing in $\bigO(m)$-time.
	It performs $5n$ text character comparisons in the worst case.}
\myitem{{AG}} {Apostolico-Giancarlo} {~\cite{AG86}} {1986} 
	{Variant of (\ref{{BM}}). It works in $\bigO(n)$-time complexity and performs $3n/2$ comparisons in the worst case.}
\myitem{{KR}} {Karp-Rabin} {~\cite{KR87}} {1987} 
	{First filter algorithm using an hashing function. It works in $\bigO(mn)$-time complexity but has an $\bigO(n+m)$ expected 	running time.}
\myitem{{ZT}} {Zhu-Takaoka} {~\cite{ZT87}} {1987} 
	{Improvement of (\ref{{HOR}}) which uses two characters for computing the occurrence heuristics.}
\myitem{{QS}} {Quick-Search} {~\cite{Sun90}} {1990} 
	{Improvement of (\ref{{HOR}}). It uses the character which follows the current window to compute the occurrence heuristics.}
\myitem{{OM}} {Optimal-Mismatch} {~\cite{Sun90}} {1990} 
	{Variation of (\ref{{QS}}) using probabilities of characters. 
	It compares the pattern and the text proceeding from the less frequent to the more frequent character. }
\myitem{{MS}} {Maximal-Shift} {~\cite{Sun90}} {1990} 
	{Variation of (\ref{{QS}}) where pattern characters are scanned from 
	the one which will lead to a larger shift to the one which will lead to a shorter shift}
\myitem{{AC}} {Apostolico-Crochemore} {~\cite{AC91}} {1991} 
	{Modification of (\ref{{KMP}}). It performs at most $3n/2$ characters inspections in the worst case.}
\myitem{{TW}} {Two-Way} {~\cite{CP91}} {1991} 
	{Linear algorithm in the worst case. Divides the pattern in two factors. It proceeds from left to right while scanning the left factor, and proceeds from right to left while scanning the right factor. It inspects at most $2n-m$ characters.}
\myitem{{TunBM}} {Tuned-Boyer-Moore} {~\cite{HS91,HS91b}} {1991}
	{Variant of (\ref{{HOR}}) which introduces a fast loop with unrolled blind shifts.}
\myitem{{COL}} {Colussi} {~\cite{Col91b}} {1991} 
	{Refinement of (\ref{{KMP}})} using a 2-factorization of the pattern.
\myitem{{SMITH}} {Smith} {~\cite{Smi91}} {1991}
	{Combination of (\ref{{HOR}}) and (\ref{{QS}}) which takes the maximum of the shifts proposed by the two occurrences heuristics.} 
\myitem{{GG}} {Galil-Giancarlo} {~\cite{GG92}} {1992} 
	{Refinement of (\ref{{COL}}) which performs at most $4n/3$ inspections in the worst case.}
\myitem{{RAITA}} {Raita} {~\cite{Rai92}} {1992} 
	{Modification of (\ref{{HOR}}). It first compare the last character, then the first, and finally the middle character. If no mismatch occurs all other characters are compared. }
\myitem{{SMOA}} {String-Matching on Ordered ALphabet} {~\cite{Cro92}} {1992} 
	{It uses constant extra space and achieves a linear time complexity. It inspects at most $6n+5$ characters.}
\myitem{{TBM}} {Turbo-Boyer-Moore} {~\cite{CCGJLPR92,CCGJLPR94}} {1992} 
	{Improvement of (\ref{{BM}}). It remembers characters inspected in the previous attempt. It performs at most $2n$ character inspections.}
\myitem{{NSN}} {Not-So-Naive} {~\cite{Han93}} {1993} 
	{Improvement of (\ref{{BF}}). It uses constant extra space and works in $\bigO(nm)$-time.}
\myitem{{RCOL}} {Reverse-Colussi} {~\cite{Col94}} {1994} 
	{Combination of (\ref{{COL}}) and (\ref{{BM}})}. It works in linear time and performs at most $2n$ character inspections.
\myitem{{SKIP}} {Skip-Search} {~\cite{CLP98}} {1998} 
	{It introduces the use of buckets of positions for each character of the alphabet. It works in $\bigO(nm)$-time but has a linear average behavior.}
\myitem{{ASKIP}} {Alpha-Skip-Search} {~\cite{CLP98}} {1998} 
	{Improvement of (\ref{{SKIP}}). It uses buckets of positions for each factor of length $\log\sigma(m)$ of the pattern.}
\myitem{{KMPS}} {Knuth-Morris-Pratt Skip-Search} {~\cite{CLP98}} {1998}
	{Improvement of (\ref{{SKIP}})} using shift tables in (\ref{{MP}}) and (\ref{{KMP}}). 
\myitem{{BR}} {Berry-Ravindran} {~\cite{BR99}} {1999} 
	{Combination of (\ref{{QS}}) and (\ref{{ZT}}). It uses the two consecutive characters, after the current window, for computing the occurrence heuristics.}
\myitem{{AKC}} {Ahmed-Kaykobad-Chowdhury} {~\cite{AKC03}} {2003} 
	{A variant of (\ref{{AG}})  that remembers all the suffixes of the pattern found in the text and that computes the shifts accordingly at the end of each attempt.}
\myitem{{FS}} {Fast-Search} {~\cite{CF03a,CF05}} {2003} 
	{Improvement of (\ref{{BM}}). It uses the occurrence heuristics when the mismatch occurs during the first comparison.}
\myitem{{FFS}} {Forward-Fast-Search} {~\cite{CF05}} {2004} 
	{Combination of (\ref{{FS}}) and (\ref{{QS}}). It computes the good-suffix using information about the character following the 	current window of the text.}
\myitem{{BFS}} {Backward-Fast-Search} {~\cite{CF05}} {2004} 
	{Combination of (\ref{{BM}}) and (\ref{{HOR}}). It computes the good-suffix heuristics using information about the mismatching 	character.}
\myitem{{TS}} {Tailed-Substring} {~\cite{CF04}} {2004} 
	{Improvement of (\ref{{BF}}) using a variant of the occurrence heuristics in (\ref{{HOR}}). It uses constant extra space.}
\myitem{{SSABS}} {Sheik-Sumit-Anindya-Balakrishnan-Sekar} {~\cite{SAPBS04}} {2004} 
	{Combination of (\ref{{QS}}) and (\ref{{RAITA}}).}
\myitem{{TVSBS}} {Thathoo-Virmani-Sai-Balakrishnan-Sekar} {~\cite{HPPC08}} {2006} 
	{Combination of (\ref{{SSABS}}) and (\ref{{BR}}).}
\myitem{{PBMH}} {Boyer-Moore-Horspool using Probabilities} {~\cite{Nebel06}} {2006} 
	{Combination of (\ref{{OM}}) and (\ref{{HOR}}).}
\myitem{{FJS}} {Franek-Jennings-Smyth} {~\cite{FJS05,FJS07}} {2007}
	{Combination of (\ref{{KMP}}) and (\ref{{QS}}).} 
\myitem{{2BLOCK}} {2-Block	Boyer-Moore	} {~\cite{SM07}} {2007} 
	{Improvement of (\ref{{BM}}) performing a constant number of inspections in the worst case.}
\myitem{{HASHq}} {Wu-Manber for Single Pattern Matching} {~\cite{Lec07}} {2007} 
	{Improvement of (\ref{{HOR}}) using a super alphabet. It computes a fingerprint of each $q$-gram in the pattern using an hash function.}
\myitem{{BMHq}} {Boyer-Moore-Horspool with $q$-grams} {~\cite{KPT08}} {2008} 
	{Improvement of (\ref{{HOR}}) using a super alphabet. It computes computes the occurrence heuristics reading $q$-grams in a single operation.}
\myitem{{TSW}} {Two Sliding Windows} {~\cite{HASIA08}} {2008} 
	{Improvement of (\ref{{QS}}) using two sliding windows. The first windows slides from left to right while the second slides from right to left.}
\myitem{{GRASPm}} {Genomic Rapid Algo for String Pm} {~\cite{DC09}} {2009} 
	{Modification of (\ref{{HOR}}). It improve the original algorithm using a filtering method based on an hash function computed on $2$-grams in the pattern.}
\myitem{{BBM}} {Bounded Boyer-Moore} {~\cite{CCF10}} {2010} 
	{Improvement of (\ref{{BM}}) using a bounded  good-suffix heuristics with constant extra space.}
\myitem{{BFS}} {Bounded Fast-Search} {~\cite{CCF10}} {2010} 
	{Improvement of (\ref{{FS}}) using a bounded  good-suffix heuristics with constant extra space.}
\myitem{{BFFS}} {Bounded Forward-Fast-Search} {~\cite{CCF10}} {2010} 
	{Improvement of (\ref{{FFS}}) using a bounded forward-good-suffix heuristics with constant extra space.}
\myitem{{FSw}} {Fast-Search using Multiple Windows} {~\cite{FL12b}} {2012} 
	{Improvement of (\ref{{FS}}) using multiple sliding windows. Windows slides from left to right and from right to left.}
\myitem{{TVSBSw}} {TVSBS using Multiple Windows} {~\cite{FL12b}} {2012} 
	{Improvement of (\ref{{TVSBS}}) using multiple sliding windows. Windows slides from left to right and from right to left.}
\myitem{{MSBM}} {Max Shift Boyer-Moore} {~\cite{SS12}} {2012}
	{Combination of (\ref{{BM}}) and (\ref{{HOR}}). It takes the maximum shift proposed by the two occurrence heuristics.} 
\myitem{{MSH}} {Max Shift Horspool} {~\cite{SS12}} {2012} 
	{Combination of (\ref{{HOR}}) and (\ref{{BM}}). It takes the maximum shift proposed by the two occurrence heuristics.} 
\myitem{{ETSW}} {Enhanced Two Sliding Windows} {~\cite{IHAAS12}} {2012} 
	{Combination of (\ref{{TSW}}) and (\ref{{BR}}).}
\myitem{{MHASHq}} {Hash$q$ using Multiple Hashing Functions} {~\cite{FL12c}} {2012} 
	{Improvement of (\ref{{HASHq}}) using multiple hash functions in order to reduce the number false positives.}
\myitem{{RSA}} {Enhanced Berry-Ravindran} {~\cite{SMS12}} {2012} 
	{Variant of (\ref{{BR}}) using four characters to compute the occurrence heuristics.}
\myitem{{ERSA}} {Enhanced RS-A} {~\cite{SHAAI}} {2013} 
	{Combination of (\ref{{RSA}}) and (\ref{{TSW}}).}
\myitem{{IOM}} {Improved Occurrence Heuristics} {~\cite{CF13,CF14a}} {2013} 
	{Improvement of (\ref{{HOR}}). It uses an improvement of the occurrence heuristics.}
\myitem{{WOM}} {Worst Occurrence Heuristics} {~\cite{CF13,CF14a}} {2013} 
	{Improvement of (\ref{{HOR}}). It computes the occurrence heuristics on the positions which leads to the average maximal shift.}
\myitem{{JOM}} {Jumping Occurrence Heuristics} {~\cite{CF13,CF14a}} {2013} 
	{Improvement of (\ref{{BR}}). It computes the occurrence heuristics on two non consecutive characters with a distance depending on the average maximal shift.}
\myitem{{SSM}} {Simple String Matching} {~\cite{AS15}} {2015} 
	{Modification of (\ref{{HOR}}). It scans the text from left to right and matches the pattern from right to left.}
\myitem{{QLQS}} {Quantum Leap Quick-Search} {~\cite{WKC15}} {2015}
	{Improvement of (\ref{{QS}}). It improves the shift performed by the occurrence heuristics by computing the  shift to left performed by the reverse of the pattern at a given fixed distance from the current window.} 
\myitem{{EERSA}} {Enhanced ERS-A} {~\cite{SIAAH15}} {2015} 
	{Combination of (\ref{{ERSA}}) and (\ref{{ETSW}}).}
\myitem{{FSW}} {Four Sliding Windows} {~\cite{HAAIS15}} {2015} 
	{Improvement of (\ref{{ERSA}}) using four sliding windows.}
\myitem{{SKIPq}} {Skip-Search using $q$-grams} {~\cite{Faro16}} {2016} 
	{Combination of (\ref{{SKIP}}) and (\ref{{HASHq}}). It computes buckets of positions for the fingerprint of each $q$-gram in the pattern.}
\end{enumerate}

\newpage

\section*{List of Algorithms based on Automata}
Also automata play a very important role in the design of efficient string matching algorithms.
For instance, the Deterministic Finite Automaton Matcher~\cite{CLRS01} was one of the first linear-time solutions, 
whereas the Backward-DAWG-Matching algorithm~\cite{CR94}  reached the optimal $\mathcal{O}(n\log_{\sigma}(m)/m)$ lower bound time complexity on the average.  
Both of them  are based on finite automata; in particular, they respectively simulate a deterministic automaton for the language
$\Sigma^{*}p$ and the deterministic suffix automaton of the reverse of $p$.

The efficiency of string matching algorithms depends on the underlying automaton used for recognizing the pattern $p$ 
and on the encoding used for simulating it. 
Here we present the list of all string matching algorithms based on deterministic automata.\\[0.2cm]

\begin{enumerate}[noitemsep]
\setcounter{enumi}{60}
\itemsep0.2cm
\myitem{{DFA}} {Deterministic-Finite-Automaton} {\cite{CLRS01}} {  no date } 
	{Linear algorithm using a deterministic finite state automaton recognizing all string whose suffix is equal to the pattern.
	Construction of the automaton can be done in $\bigO(m)$-time.}
\myitem{{RF}} {Reverse-Factor} {\cite{Lec92}} {1992} 
	{Combination of (\ref{{DFA}}) and (\ref{{BM}}). It uses  the suffix automaton of the reverse of the pattern. It works in $\bigO(nm)$ worst case time.}
\myitem{{SIM}} {Simon} {\cite{Sim94}} {1994} 
	{Modification of (\ref{{DFA}}). It uses the minimal automaton of the pattern.}
\myitem{{TRF}} {Turbo-Reverse-Factor} {\cite{CCGJLPR94}} {1994} 
	{Refinement of (\ref{{RF}}). It remembers the characters matched in the previuos attempt. It's complexity id $\bigO(n)$ in time.}
\myitem{{FDM}} {Forward-DAWG-Matching} {\cite{CR94}} {1994} 
	{It is a linear algorithm using the suffix automaton of the pattern.}
\myitem{{BDM}} {Backward-DAWG-Matching} {\cite{CR94}} {1994} 
	{Variant of (\ref{{RF}}). It uses the Directed Acyclic Word Graph of the pattern.}
\myitem{{BOM}} {Backward-Oracle-Matching} {\cite{ACR99}} {1999} 
	{Variant of (\ref{{RF}}). It uses the Factor Oracle of the pattern. It is the first filtering algorithm using automata.}
\newpage
\myitem{{DFDM}} {Double Forward DAWG Matching} {\cite{AR00,AR00b}} {2000} 
	{Modification of (\ref{{FDM}}). It works in linear worst case time using a DAWG of the pattern and a DAWG of the reverse of the pattern.}
\myitem{{WW}} {Wide Window} {\cite{HFS05}} {2005} 
	{Combination of (\ref{{FDM}}) and (\ref{{RF}}). It uses the suffix automaton of the pattern and the prefix automaton of the reverse of the pattern. It has a linear worst case time complexity.}
\myitem{{LDM}} {Linear DAWG Matching} {\cite{HFS05}} {2005} 
	{Combination of (\ref{{BDM}}) and (\ref{{DFA}}). It uses the suffix automaton of the reverse of the pattern deterministic finite state automaton of the pattern.}
\myitem{{ILDM1}} {Improved Linear DAWG Matching} {\cite{LWLL06}} {2006} 
	{Improvement of (\ref{{LDM}}).}
\myitem{{ILDM2}} {Improved Linear DAWG Matching 2} {\cite{LWLL06}} {2006} 
	{Improvement of (\ref{{LDM}}).}
\myitem{{EBOM}} {Extended Backward Oracle Matching} {\cite{FL08,FL09}} {2008} 
	{Improvement of (\ref{{BOM}}). It performs transitions of two characters at each step.}
\myitem{{FBOM}} {Forward Backward Oracle Matching} {\cite{FL08,FL09}} {2008} 
	{Combination of (\ref{{EBOM}}) and (\ref{{QS}}). It performs transitions using the characters which follows the current window of the text.}
\myitem{{SBDM}} {Succint Backward DAWG Matching} {\cite{Fred08}} {2008} 
	{Variant of (\ref{{BDM}}). It is based on the combination of compressed self-indexes and BDM.}
\myitem{{SEBOM}} {Simplified Extended Backward Oracle Matching} {\cite{FYM09}} {2009} 
	{Simplification of (\ref{{EBOM}}).}
\myitem{{SFBOM}} {Simplified Forward Backward Oracle Matching} {\cite{FYM09}} {2009} 
	{Simplification of (\ref{{FBOM}}).}
\myitem{{BSDM}} {Backward SNR DAWG Matching} {\cite{FL12}} {2012}
	{Modification of (\ref{{BDM}}). It uses the DAWG constructed on the longest substring oft he pattern with no repetitions of characters.} 
\myitem{{BSDMqx}} {BSDM using $q$-grams and shift-xor} {\cite{Faro16b}} {2016} 
	{Variant of (\ref{{BSDM}}). It computes the DAWG assuming a super alphabet on $q$-grams of the pattern. The $q$-grams are associated to a fingerprint value computed by using a shift-xor hash function.}
\myitem{{BSDMqxw}} {BSDM$qx$ using multiple windows} {\cite{Faro16b}} {2016} 
	{Improvement of (\ref{{BSDMqx}}) which uses multiple sliding windows.}
\end{enumerate}

\section*{List of Algorithms based on Bit Parallelism}
Bit-parallelism  \cite{BYG89,BYG92} is a technique used for simulating nondeterministic automata.
Specifically the bit-parallelism technique takes advantage of the intrinsic parallelism of
the bitwise operations inside a computer word, allowing to cut down the
number of operations that an algorithm performs by a factor up to $w$,
where $w$ is the number of bits in the computer word.
However the correspondent encoding requires one bit per pattern symbol, for a
total of $\mw$ computer words.  Thus, as long as a pattern fits in a
computer word, bit-parallel algorithms are extremely fast, otherwise
their performances degrades considerably as $\mw$ grows. 

Here is the list of all string matching algorithms based on bit-parallelism.

\begin{enumerate}[noitemsep]
\setcounter{enumi}{80}
\itemsep0.2cm
\myitem{{SO}} {Shift-Or} {\cite{BYG89,BYG92}} {1989} 
	{Simulates the nondeterministic version of the automaton in (\ref{{DFA}}).}
\myitem{{SA}} {Shift-And} {\cite{BYG89,BYG92}} {1989} 
	{Simulates the nondeterministic version of the automaton in (\ref{{DFA}}).}
\myitem{{BNDM}} {Backward-Nondeterministic-DAWG-Matching} {\cite{NR98a}} {1998} 
	{Simulates the nondeterministic version of the automaton in (\ref{{BDM}}).}
\myitem{{BNDML}} {BNDM for Long patterns} {\cite{NR00}} {2000 } 
	{Modification of (\ref{{BNDM}}) using multiple words for simulating the nondeterministic automata of long patterns.}
\myitem{{SBNDM}} {Simplified BNDM} {\cite{PT03,Nav01}} {2003 } 
	{Improvement of (\ref{{BNDM}}).}
\myitem{{TNDM}} {Two-Way Nondeterministic DAWG Matching} {\cite{PT03}} {2003} 
	{A two-way modification of (\ref{{BNDM}}). It scans a pattern suffix forward before normal backward scan.}
\myitem{{LBNDM}} {Long patterns BNDM} {\cite{PT03}} {2003 } 
	{Modification of  (\ref{{BNDM}}) for long patterns. The automaton is constructed over a superimposed pattern constructed on the $m/w$ consecutive factors of the inout pattern.}
\myitem{{SVM}} {Shift Vector Matching} {\cite{PT03}} {2003 } 
	{Implementation of (\ref{{BM}}) using bit parallelism. It remembers characters matched during the last attempt.}
\myitem{{BNDM2}} {BNDM with loop-unrolling} {\cite{HD05}} {2005 } 
	{Improvement of (\ref{{BNDM}}) using a blind enrolled loop.}
\myitem{{SBNDM2}} {Simplified BNDM with loop-unrolling} {\cite{HD05}} {2005 } 
	{Improvement of (\ref{{SBNDM}}) using a blind enrolled loop.}
\myitem{{BNDMBMH}} {BNDM with Horspool Shift} {\cite{HD05}} {2005 } 
	{Combination of (\ref{{BNDM}}) and (\ref{{HOR}})}
\myitem{{BMHBNDM}} {Horspool with BNDM test} {\cite{HD05}} {2005 } 
	{Combination of (\ref{{BNDM}}) and (\ref{{HOR}})}
\myitem{{FNDM}} {Forward Nondeterministic DAWG Matching} {\cite{HD05}} {2005 } 
	{Simulates the nondeterministic version of the automaton in (\ref{{FDM}}).}
\myitem{{BWW}} {Bit parallel Wide Window} {\cite{HFS05}} {2005 } 
	{Simulates the nondeterministic version of the automaton in (\ref{{WW}}).}
\myitem{{AOSO}} {Average Optimal Shift-Or} {\cite{FG05}} {2005 } 
	{Variant of (\ref{{SO}}) using a superimposed pattern. It improves the original algorithm by shifting of more than one position during the scan.}
\myitem{{FAOSO}} {Fast Average Optimal Shift-Or} {\cite{FG05}} {2005 } 
	{Practical improvement of (\ref{{AOSO}}).}
\myitem{{FBNDM}} {Forward BNDM} {\cite{FL08,FL09}} {2008} 
	{Simulates the nondeterministic version of the automaton in (\ref{{FBOM}}).}
\myitem{{FSBNDM}} {Forward Simplified BNDM} {\cite{FL08,FL09}} {2008 } 
	{Combination of (\ref{{SBNDM}}) and (\ref{{FBNDM}}).}
\myitem{{BLIM}} {Bit-Parallel Length Invariant Matcher} {\cite{OK08}} {2008 } 
	{Quadratic worst case time algorithm using multiple words.}
\myitem{{BNDMq}} {BNDM with $q$-grams} {\cite{DHPT09}} {2009 } 
	{Variant of (\ref{{BNDM}}) using a super alphabet implemented with $q$-grams.}
\myitem{{SBNDMq}} {Simplified BNDM with $q$-grams} {\cite{DHPT09}} {2009 } 
	{Variant of (\ref{{SBNDM}}) using a super alphabet implemented with $q$-grams.}
\myitem{{UFNDMq}} {FNDM with $q$-grams} {\cite{DHPT09}} {2009 } 
	{Implementation of (\ref{{FNDM}}) using $q$-grams.}
\myitem{{SABP}} {Small Alphabet Bit-Parallel} {\cite{ZZMY09}} {2009 } 
	{Bit Parallel algorithm designed for searching string over small alphabets. It is based on a position related character matching table.}
\myitem{{BXS}} {BNDM with Extended Shifts} {\cite{DPST10}} {2010} 
	{Modification of  (\ref{{BNDM}}) for long patterns. The automaton is constructed over a superimposed pattern constructed on the $m/w$ consecutive factors of the input pattern.}
\myitem{{BQL}} {BNDMq Long} {\cite{DPST10}} {2010} 
	{Modification of  (\ref{{BNDM}}) for long patterns. It increases the effective alphabet size by using overlapping $q$-grams.}
\myitem{{QF}} {Q-Gram Filtering} {\cite{DPST10}} {2010} 
	{It uses a filtering approach based on consecutive $q$-grams in the text.}
\myitem{{BP2WW}} {\bpdww} {\cite{CFG10}} {2010 } 
	{Improvement of (\ref{{WW}}) using two sliding windows inspected in parallel.}
\myitem{{BPWW2}} {\bpwwd} {\cite{CFG10}} {2010 } 
	{Improvement of (\ref{{WW}}) using two sliding windows inspected in parallel.}
\myitem{{KSA}} {Factorized Shift-And} {\cite{CFG10a,CFG12}} {2010 } 
	{Simulates the nondeterministic version of the automaton in (\ref{{DFA}}) using a more compact representation of the automaton. It is based on a factorization of the pattern.}
\myitem{{KBNDM}} {Factorized BNDM} {\cite{CFG10a,CFG12}} {2010 } 
	{Simulates the nondeterministic version of the automaton in (\ref{{BNDM}}) using a more compact representation of the automaton. It is based on a factorization of the pattern.}
\myitem{{FSBNDMqf}} {Forward SBNDM using $q$-grams and Lookahead} {\cite{PT11,PT14}} {2011 } 
	{Improvement of (\ref{{FSBNDM}}) which implements a super alphabet with $q$-grams and computing the shift with a lookahead of several characters.}
\myitem{{SBNDMw}} {SBNDM using Multiple Windows} {\cite{FL12b}} {2012 } 
	{Improvement of (\ref{{SBNDM}}) using multiple sliding windows.}
\myitem{{FSBNDMw}} {Forward SBNDM using Multiple Windows} {\cite{FL12b}} {2012 } 
	{Improvement of (\ref{{FSBNDM}}) using multiple sliding windows.}
\myitem{{TSA}} {Improved Two-Way Shift-And} {\cite{DCGHPT14}} {2014 } 
	{Improvement of (\ref{{SA}}) using several sliding windows inspected i parallel.}
\myitem{{TSO}} {Improved Two-Way Shift-Or} {\cite{DCGHPT14}} {2014 } 
	{Improvement of (\ref{{SO}}) using several sliding windows inspected i parallel.}
\myitem{{TSAq}} {Two-Way Shift-And using $q$-grams} {\cite{DCGHPT14}} {2014 } 
	{Improvement of (\ref{{SA}}) using several sliding windows inspected i parallel. It improves the performances by simulating a super alphabet with $q$-grams}
\myitem{{TSOq}} {Two-Way Shift-Or using $q$-grams} {\cite{DCGHPT14}} {2014 } 
	{Improvement of (\ref{{SO}}) using several sliding windows inspected i parallel. It improves the performances by simulating a super alphabet with $q$-grams}
\end{enumerate}

\newpage
\section*{List of Algorithms based on Packed String Matching}
In the packed string matching technique multiple characters are packed into one larger word, so that the
characters can be compared in bulk rather than individually. In this context, if the characters of a string are drawn
from an alphabet of size $\sigma$, then $\lfloor \frac{w}{\log \sigma} \rfloor$ different characters fit in a single word, using
$\lceil \log \sigma \rceil$ bits per characters. The packing factor is $\alpha= \lfloor \frac{w}{\log \sigma} \rfloor$. 

\begin{enumerate}[noitemsep]
\setcounter{enumi}{117}
\itemsep0.2cm
\myitem{{SAS}} {Super Alphabet Simulation} {\cite{Fred02}} {2002} 
	{General approach for packed string matching based on a tabulation technique.}
\myitem{{PSS}} {Packed String Search} {\cite{Bille09,Bille11}} {2009}
	{Variant of (\ref{{KMP}}) applied to packed string matching.} 
\myitem{{SSEF}} {Streaming SIMD Extensions Filter} {\cite{KO09}} {2009} 
	{Extension of (\ref{{SKIP}}) to packed string matching. It is implemented using SSE instructions. It works only for long patterns.}
\myitem{{PB}} {Packed Belazzougui} {\cite{Belaz10,Belaz12}} {2010} 
	{Efficient packed string matching algorithm which works in $\bigO(n/\alpha + occ)$-time when $\alpha \leq m\leq n/\alpha$.}
\myitem{{PBR}} {Packed Belazzougui-Raffinot} {\cite{BR13}} {2010} 
	{Efficient packed string matching algorithm which works in $\bigO(nm)$ worst case time complexity.}
\myitem{{SSECP}} {Crochemore-Perrin algorithm using SSE instructions} {\cite{BBBGGW11,BBBGGW14}} {2011} 
	{Modification of (\ref{{TW}}) for packed string matching. It is based on two specialized packed string instructions.}
\myitem{{EPSM}} {Exact Packed String Matching} {\cite{FK13,FK14}} {2013} 
	{Improvement of (\ref{{SSEF}}). It uses SSE instruction to speed up searching. It consists in four different algorithms depending on the length of the pattern.}
\end{enumerate}

\newpage

\bibliographystyle{splncs03}
\bibliography{biblio}

\end{document}